\documentclass[10pt, journal]{IEEEtran}
\ifCLASSINFOpdf
\else
\fi

\usepackage{graphicx}
\usepackage{epstopdf}
\usepackage{amsfonts,amsmath,amssymb,hyperref, amsthm}
\usepackage{cite}
\usepackage{dsfont}
\usepackage{bm}
\usepackage{amssymb}
\usepackage{array}
\usepackage{algorithm}
\usepackage{algorithmic}
\usepackage{subfigure}
\usepackage{graphics}
\usepackage{epsfig}
\usepackage{float}
\usepackage{extarrows}
\usepackage{xcolor}
\usepackage{verbatim}

\newtheorem{theorem}{Theorem}
\newtheorem{corollary}{Corollary}

\theoremstyle{definition}

\begin{document}
%\abovedisplayskip=1pt
%\belowdisplayskip=1pt
%\allowdisplaybreaks

\title{Double-Sided Information Aided Temporal-Correlated Massive Access}

%\author{
%\IEEEauthorblockN{Weifeng~Zhu, Meixia~Tao, and Yunfeng~Guan}
%
%\IEEEauthorblockA{Department of Electronic Engineering, Shanghai Jiao Tong University, Shanghai, China \\
%Emails: \{wf.zhu, mxtao, yfguan69\}@sjtu.edu.cn} \\
%}

\author{
Weifeng~Zhu,~\IEEEmembership{Graduate Student Member,~IEEE}, Meixia~Tao,~\IEEEmembership{Fellow,~IEEE}, and Yunfeng~Guan

%%\thanks{This paper was presented in part at the IEEE International Conference of Communications (ICC) 2021 \cite{Zhu_2021_ICC}.}
\thanks{The authors are with the Department of Electronic Engineering, Shanghai Jiao Tong University, Shanghai 200240,
China (e-mail: \{wf.zhu, mxtao, yfguan69\}@sjtu.edu.cn).}
\vspace{-1cm}

%%
%%%\thanks{X. Yuan is with the Center for Intelligent Networking and Communication (CINC), University of Electronic Science and Technology of China, Chengdu 610000, China (e-mail: xjyuan@uestc.edu.cn).}
%%%\thanks{ is with the Department of Electronic Engineering, Shanghai Jiao Tong University, Shanghai 200240,
%%%China  (e-mail: @sjtu.edu.cn).}
}

%\vspace{-0.8cm}

%\thanks{This paper was presented in part at the IEEE International Conference of Communications (ICC) 2021 \cite{Zhu_2021_ICC}.}
%\thanks{This work is supported by the National Key R\&D Project of China under grant 2020YFB1406802 and the NSF of China under grant 61941106. }
%}

%\markboth{IEEE Wireless Communication Letters,~Vol.~xx, No.~xx}{W. Zhu \MakeLowercase{\textit{et al.}}: Generalized Side Information Aided Temporal-Correlated Massive Access}

\maketitle

\begin{abstract}

This letter considers temporal-correlated massive access, where each device, once activated, is likely to transmit continuously over several consecutive frames. Motivated by that the device activity at each frame is correlated to not only its previous frame but also its next frame, we propose a double-sided information (DSI) aided joint activity detection and channel estimation algorithm based on the approximate message passing (AMP) framework. The DSI is extracted from the estimation results in a sliding window that contains the target detection frame and its previous and next frames. The proposed algorithm demonstrates superior performance over the state-of-the-art methods.

%This paper considers joint activity detection and channel estimation problem in the temporal-correlated massive access systems, where massive devices are sporadically activated for information transmission. The temporal-correlated device activity demonstrates that the active users in the current frame are highly possible to keep transmitting information in the following frames. By accounting both the sparsity and temporal correlation in the device activity, the joint activity detection and channel estimation problem can be formulated as a dynamic compressed sensing (DCS) problem.
%However, the conventional approaches usually give a sub-optimal solution to the DCS problem, since they only utilize part of the available side information (SI) containing the estimated activity information in the previous frame.
%By leveraging approximate message passing with SI (AMP-SI), two generalized SI-aided joint activity detection and channel estimation frameworks are proposed to fully exploit the temporal correlation for performance enhancement, which can take the SI from any number of neighboring frames into consideration. Moreover, the generalized SI-aided device activity detector is also designed based on the log likelihood ratio (LLR) test. The simulation results demonstrate that the proposed algorithm can realize significantly superior performance to the benchmarks.

\end{abstract}

\begin{IEEEkeywords}
Massive access, temporal correlation, approximate message passing (AMP), side information.
\end{IEEEkeywords}

\section{Introduction}

Massive machine-type communication (mMTC) is one of the main use cases of the 5G and beyond mobile communication systems for supporting Internet of Things (IoT) applications. It is featured by the massive number of IoT devices and their sporadic activities \cite{Liu_2018_SPM}. To enable massive connectivity, grant-free random access is introduced in 5G as a new random access technique where devices can access channel resources without undergoing a handshake process. Thereby, the signaling overhead can be efficiently reduced. However, the main technical challenge of grant-free random access is device activity detection and channel estimation.

%A main technical challenge in grant-free random access is activity detection and channel estimation.
%The grant-free random access schemes usually pre-allocate a unique pilot sequence to each devices for user identification and channel estimation. However, the limited time-frequency resource usually leads to the employment of non-orthogonal pilot sequences, which causes significant co-channel interference.
By exploiting the sporadic nature of data traffic in mMTC, compressed sensing (CS) techniques have been widely utilized to perform activity detection and channel estimation jointly in the literature \cite{Liu_2018_SPM}. Among all these CS-based methods, the approximate message passing (AMP) based algorithms have demonstrated favorable performance at efficient computation complexity in \cite{Chen_2018_TSP, Liu_2018_TSP, Senel_2018_TCOM, Ke_2020_TSP}.
The covariance-based method \cite{Fengler_2021_TIT} is another popular approach for activity detection, which can outperform the conventional CS-based methods when the number of antennas is large.
%\textcolor[rgb]{0.00,0.07,1.00}{When the BS is equipped with a massive number of antennas, the covariance-based algorithm proposed in \cite{Fengler_2021_TIT} can detect much more active devices than the traditional CS-based methods with equal pilot length.}

In practical IoT environment, if a device is activated by a burst event, the device often transmits continuously for a certain time interval. In other words, once activated, it often remains active in the upcoming consecutive transmission frames. This suggests that the device activity is correlated in the time domain. By accounting such temporal correlation, the performance of activity detection and channel estimation can be improved by formulating the problem from the dynamic CS (DCS) perspective \cite{Jiang_2021_TWC, Zhu_2021_ICC, Wang_2021_ISIT}. Specifically, the work \cite{Jiang_2021_TWC} proposes a sequential AMP (S-AMP) algorithm for device activity detection by using the historical knowledge. In \cite{Zhu_2021_ICC}, a hybrid generalized AMP (HyGAMP) algorithm is employed to account the device activity information from both the previous frame and the next frame to improve performance. Note that both \cite{Jiang_2021_TWC} and \cite{Zhu_2021_ICC} focus on the scenario where the base station (BS) has single antenna. In \cite{Wang_2021_ISIT}, a side information aided multiple measurement vector based AMP (SI-aided MMV-AMP) algorithm is proposed for both single-antenna and multiple-antenna scenarios. Here, the SI is defined as the estimation result in the previous frame. However, the algorithm only considers the SI from the single previous frame and thus the temporal correlation is not fully exploited. %Thus, the performance improvement is limited since not all the available SI is utilized in the detection and estimation algorithm.
%However, the temporal correlations are not fully exploited in the SIA-MMV-AMP algorithm, which limits the detection and estimation performance.

This work aims to fully exploit the temporal correlation of device activity for further enhancing the joint activity detection and channel estimation performance. This is motivated by the fact that the activity of each device in a current frame is not only related to the activity in the previous frame (single side) but also correlated with the activity in the next frame (double sides). More specifically, if we know one device is active in the previous frame, then with a large probability it will be considered to be active in the current frame; but if we know that the device is active in both the previous and next frames, almost surely it is active in the current frame. Therefore, by exploiting the estimation results from the adjacent frames in double sides, one can further lift the performance of activity detection and channel estimation at each current frame.

To this end, we first introduce a sliding-window detection strategy to account the double-sided information (DSI) from the previous frame as well as the next frame.
%Firstly, we point out that the performance can be greatly improved by additionally accounting the SI from the next frame.
Then we propose a DSI-aided activity detection and channel estimation algorithm based on the AMP-SI framework proposed in \cite{Ma_2019_TSP}. The proposed activity detector employs the log likelihood ratio (LLR) test. %We also show that the proposed algorithm can be extend to the generalized case where the activity evolution satisfies more complex stochastic process.
Numerical results show that the proposed algorithm significantly outperforms the state-of-the-art algorithms by further exploiting the temporal correlation of device activity.

\section{System Model}

Consider a grant-free massive access system, where a very large number $N$ of single-antenna user devices communicate with a common $M$-antenna BS through a shared uplink channel. Due to the sporadic communication traffic, only a small subset of devices are activated in each transmission frame. Each  frame consists of two phases, a pilot phase and a data phase. We concentrate on the pilot phase for activity detection and channel estimation. A unique pilot sequence $\mathbf{a}_{n} = [a_{n,1}, a_{n,2}, \dots, a_{n,L}]^T \in \mathbb{C}^{L \times 1}$ is pre-allocated to each device $n$, $\forall n \in \{1,\dots,N\}$, for identification and channel estimation, where $L \ll N$ is the pilot length. We assume that the elements of each pilot sequence are generated following the independent and identically distributed (i.i.d.) complex Gaussian distribution with zero mean and variance $\frac{1}{L}$, i.e., $a_{n,l} \sim \mathcal{CN}(0,\frac{1}{L}), \forall n,l$.

\subsection{Temporal-Correlated Activity Model}\label{subsec:TC_Activity}

We consider temporally correlated device activity.
Let $\lambda_n^t \in \{0,1\}$ denote the activity state of device $n$ at the $t$th frame, with $\lambda_{n}^t=1$ meaning active and $\lambda_{n}^t=0$ otherwise. Without loss of generality, we assume $\lambda_n^t$ evolves over $t$ according to a stationary stochastic process and is i.i.d. for all devices. As in \cite{Jiang_2021_TWC, Zhu_2021_ICC, Wang_2021_ISIT}, the device activity evolution is modeled by a first-order steady Markov chain, which can be fully described by two transition probabilities $p_{01} = \text{Pr}(\lambda_n^{t+1}=1|\lambda_n^t=0)$ and  $p_{11} = \text{Pr}(\lambda_n^{t+1}=1|\lambda_n^t=1)$.
%Due to the fact that the sum of the elements in each column of $\mathbf{P}$ equals one, the Markov chain can be completely described by the two parameters $p_{01}$ and $p_{11}$.
Then the active probability of each device, denoted as $p_a$, in each frame can be derived by solving the eigenvalue problem as $p_a = \frac{p_{01}}{1-p_{11}+p_{01}}$.
%\begin{align}\label{equ:pa}
%  p_a &= \frac{p_{01}}{1-p_{11}+p_{01}}.
%\end{align}
Note that $p_{01}$ and $p_{11}$ can be estimated empirically based on the historical data. In the special case where $p_{01} = p_{11}$, the device activity is independent over time.
%In fact, the statistical relations between the user activities in the adjacent frames are usually unknown, but we find that the above model is precise enough for us to exploit the temporal correlation in the detection and estimation procedure\footnote{The parameters $p_{01}$ and $p_{11}$ in the model can be learned from the pre-collected data by the machine learning methods, e.g., the EM algorithm.}.

\vspace{-0.5cm}
\subsection{Signal Model}
We assume block-fading channel model where the channel of each device remains unchanged in one frame, but varies in different frames. In the $t$th frame, the channel
coefficient vector between the BS and device $n$ is defined as $\mathbf{h}_{n}^{t} = \sqrt{\beta_n} \mathbf{g}_n^t \in \mathbb{C}^{1 \times M}$, where $\beta_n = \rho_n \gamma_n \in \mathbb{R}$ represents the large-scale fading coefficient affected by the attenuation factor $\gamma_n$ and the transmit power $\rho_n$, $\mathbf{g}_n^t \in \mathbb{C}^{1 \times M}$ is the small-scale fading vector. This work adopts the simple power control strategy in \cite{Senel_2018_TCOM} based on the attenuation factors to benefit cell-edge devices. Thus, we have $\beta_n = \beta, \forall n$. The small-scale fading is assumed to follow i.i.d. Rayleigh distribution, i.e., $\mathbf{g}_n^t \sim \mathcal{CN}(0,\mathbf{I}), \forall n,t$. Then during the pilot phase, the received signals at the BS in the $t$th frame, denoted as $\mathbf{Y}^{t} \in \mathbb{C}^{L \times M}$, can be written as
\begin{align}\label{equ:Y}
    \mathbf{Y}^t %&= \sum_{n=1}^{N} \lambda^{t}_{n} \mathbf{a}_{n} \mathbf{h}_n^t + \mathbf{W}^t,  \notag \\
    &= \mathbf{A} \pmb{\Lambda}^t \mathbf{H}^t + \mathbf{W}^t = \mathbf{A} \mathbf{X}^t + \mathbf{W}^t,
\end{align}
where $\mathbf{A} = [\mathbf{a}_1, \mathbf{a}_2, \dots, \mathbf{a}_N] \in \mathbb{C}^{L \times N}$ is the pilot matrix; $\pmb{\Lambda}^t = \text{diag}([\lambda_1^t, \lambda_2^t, \dots, \lambda_N^t]^T) \in \mathbb{C}^{N \times N}$ is the activity matrix; $\mathbf{H}^t = [(\mathbf{h}_1^t)^T, (\mathbf{h}_2^t)^T, \dots, (\mathbf{h}_N^t)^T]^T \in \mathbb{C}^{N \times M}$ is the channel matrix; $\mathbf{X}^t = \pmb{\Lambda}^t \mathbf{H}^t \in \mathbb{C}^{N \times M}$ represents the effective channel matrix; $\mathbf{W}^t \in \mathbb{C}^{L \times M}$ is the additive noise matrix whose elements satisfy i.i.d. complex Gaussian distribution with zero mean and variance $\sigma^2_w$.

To perform activity detection and channel estimation is essential to recover $\mathbf{X}^t$ from $\mathbf{Y}^t$. The problem can usually solved by the AMP algorithm with vector shrinkage function (vAMP) \cite{Chen_2018_TSP, Liu_2018_TSP}. This approach, however, treat $\mathbf{X}^{t}$ in different frames independently and thus ignores the temporal correlation. In the following, we propose to leverage the temporal correlation of device activity by exploiting the double-sided information under the vAMP framework.

%With the received signal $\mathbf{Y}^t$ and known pilot matrix $\mathbf{A}$, the task is to recover the effective channel matrix $\mathbf{X}^t$, which is then utilized for active user detection. Such problem is usually formulated as a CS-MMV problem, where the BS treats $\mathbf{X}^t$ in different frames independently. This approach, however, ignores the temporal correlation. Therefore, we propose to reformulate the problem from the DCS-MMV perspective, which can account the relationship between the user activities in all adjacent frames to enhance the detection and estimation performance.

\section{vAMP Framework with Side Information}

\subsection{Overview of the SI-Aided vAMP}
In this subsection, we first introduce the SI-aided vAMP framework, denoted as vAMP-SI. The general procedure of the framework roots from the AMP algorithm with SI proposed in \cite{Ma_2019_TSP} and operates as follows.
Let $\widehat{\mathbf{X}}^t_i$ denote the estimation of the effective channel $\mathbf{X}^t$ in the $i$th iteration. Also, let $\mathbf{V}^t_{i}$ denote the residual of the received signal $\mathbf{Y}^t$ corresponding to the estimation $\widehat{\mathbf{X}}^t_i$.
Starting with $\widehat{\mathbf{X}}^t_0 = \bf{0}$ and $\mathbf{V}^t_0 = \mathbf{Y}^t$, the algorithm computes at the $i$th iteration:
\begin{align}
    \mathbf{R}^t_{i} =&~ \widehat{\mathbf{X}}^t_{i-1} + \mathbf{A}^H \mathbf{V}^t_{i-1},  \\
    \widehat{\mathbf{X}}^t_{i} =&~ \pmb{\eta}_{i}\left(\mathbf{R}^t_{i},\mathcal{S}^t\right),\\
    \mathbf{V}^t_{i} =&~ \mathbf{Y}^t - \mathbf{A}\widehat{\mathbf{X}}^t_{i} + \frac{1}{L}\mathbf{V}^t_{i-1}\sum_{n=1}^{N} \frac{\partial \pmb{\eta}_{n,i}(\mathbf{r}^t_{n,i},\mathcal{S}^{t}_n)}{\partial \mathbf{r}^{t}_{n,i}}, \label{equ:Vt}
 \end{align}
where
%$\widehat{\mathbf{X}}_{i}^t \in \mathbb{C}^{N \times M}$ is the estimate of the effective channel matrix $\mathbf{X}^{t}$ at the $i$th iteration; $\mathbf{V}^{t}_{i} \in \mathbb{C}^{L \times M}$ is the residual of the received signal $\mathbf{Y}^{t}$;
$\mathbf{R}^{t}_{i} \in \mathbb{C}^{N \times M}$ can be viewed as the matched filtered output on the iteration-$(i-1)$ residual measurement $\mathbf{V}^t_{i-1}$;
%$\Sigma^{t}_{i}$ is the covariance matrix of shrinkage input $\mathbf{R}^{t}_{i}$ which can be modeled as an additive white Gaussian noise-corrupted version of the true signal $\mathbf{X}^{t}_{0}$;
%$\mathbf{S}^{t} = [(\mathbf{s}_{1}^{t})^T,\dots,(\mathbf{s}_{N}^{t})^T]^T$ with each $\mathbf{s}^{t}_{n}$ representing the SI for device $n$ in the $t$th frame;
$\mathcal{S}^t = \{\mathcal{S}^{t}_{n}\}_{n=1}^{N}$ with each $\mathcal{S}^{t}_{n}$ being the SI to improve the recovery quality of $\mathbf{x}^t_n$ for each device $n$;
$\pmb{\eta}_{i}(\cdot,\mathcal{S}^{t}) = \left[(\pmb{\eta}_{1,i}(\cdot,\mathcal{S}^{t}_{1}))^T,\dots,(\pmb{\eta}_{N,i}(\cdot,\mathcal{S}^{t}_{N}))^T\right]^T$ with each $\pmb{\eta}_{n,i}(\cdot,\mathcal{S}^{t}_{n}): \mathbb{C}^{1 \times M} \to \mathbb{C}^{1 \times M}$ being the SI-aware shrinkage function that operates on the $n$th row of $\mathbf{R}^{t}_{i}$, denoted as $\mathbf{r}^{t}_{n,i}$.
Note that the residual calculation in (\ref{equ:Vt}) also includes the ``Onsager correction'' term $\frac{1}{L}\mathbf{V}^t_{i-1}\sum_{n=1}^{N} \frac{\partial \pmb{\eta}_{n,i}(\mathbf{r}^t_{n,i},\mathcal{S}^{t}_n)}{\partial \mathbf{r}^{t}_{n,i}}$.

The key difference between vAMP-SI and vAMP lies in the introduction of SI in the shrinkage function $\pmb{\eta}_{n,i}(\cdot,\mathcal{S}^t_{n})$.
Thus, the definition of SI plays a vital role in the vAMP-SI framework.
According to the activity model in Section \ref{subsec:TC_Activity}, the activity of each device $n$ in the current frame is correlated to those in the previous and next frames. This motivates us to extract the SI from the estimation results in these two frames.
%$\mathcal{R}^{t} = \{\widehat{\mathbf{R}}^{t'}_{\infty}\}_{t' \in \mathcal{T}_w}$ and $\mathcal{R}^{t}_{n} = \{\widehat{\mathbf{r}}^{t'}_{n,\infty}\}_{t' \in \mathcal{T}_w}$ are defined as the set containing the estimation results of the frame in $\mathcal{T}_w$;

\subsection{SI Acquisition}\label{subsec:SI_Ac}
Similar to \cite{Wang_2021_ISIT}, the considered SI in this paper is identified from the state evolution of vAMP-SI in the asymptotic regime where $L, N \to \infty$ with fixed $L/N$ and $p_a$.
In specific, at the estimation of $\mathbf{x}^t_n$ for each device $n$ in each $i$th iteration, the matched filtered output $\mathbf{r}^{t}_{n,i}$ in the AMP-based algorithm can be accurately modeled by
\begin{align}\label{equ:r_model}
    \mathbf{r}^{t}_{n,i} &= \mathbf{x}^{t}_{n} + \mathbf{d}^{t}_{n,i}(\pmb{\Sigma}^t_{i-1})^{\frac{1}{2}},
\end{align}
where $\mathbf{x}^{t}_{n} \in \mathbb{C}^{1 \times M}$ is the $n$th row of $\mathbf{X}^{t}$, $\mathbf{d}^{t}_{n,i} \in \mathbb{C}^{1 \times M}$ is the corrupting noise whose elements follow i.i.d. $\mathcal{CN}(0,1)$, $\pmb{\Sigma}^t_i \in \mathbb{C}^{M \times M}$ is the covariance matrix of $\mathbf{r}^{t}_{n,i}$. Here, $\pmb{\Sigma}^{t}_{i}$ is also known as the \emph{state} of vAMP-SI in the $i$th iteration and it evolves as
\begin{align}\label{equ:SE}
    \pmb{\Sigma}_{i}^t = &~ \sigma^2_w \mathbf{I} + \frac{N}{L} \mathbb{E}\left[(\mathbf{q}^t_{n,i-1})^H \mathbf{q}^t_{n,i-1}\right],
\end{align}
where $\mathbf{q}^t_{n,i-1} =  \pmb{\eta}_{n,i-1} \left(\mathbf{x}^t_{n}+\mathbf{d}_{n,i}^t(\pmb{\Sigma}^t_{i-1})^{\frac{1}{2}},\mathcal{S}^{t}_{n}\right) - \mathbf{x}^t_{n}$.
%with random vector $\bm{U}^{t}_{n} \in \mathbb{C}^{1 \times M}$ following the distribution of $\mathbf{x}^{t}_{n}$ and random vector $\bm{E}^{t}_{n} \in \mathbb{C}^{1 \times M}$ following $\mathcal{CN}(0,\mathbf{I})$.
Here, the expectation is performed over both $\mathbf{x}^{t}_{n}$ and $\mathbf{d}^{t}_{n,i}$.

Let $\mathbf{R}^{t}_{\infty} = [\mathbf{r}^{t}_{1,\infty})^T,\dots,(\mathbf{r}^{t}_{N,\infty})^T]^T$ and $\pmb{\Sigma}^{t}_{\infty}$ denote the converged matched filtered output and the converged state in vAMP-SI in each frame $t$. According to (\ref{equ:r_model}), the statistical relationship between $\mathbf{r}^{t}_{n,\infty}$ and $\mathbf{x}^{t}_{n}$ can be modeled as
\begin{align}\label{equ:rx_infty}
    \mathbf{r}^{t}_{n,\infty} = \mathbf{x}^{t}_{n} + \mathbf{d}^{t}_{n,\infty}(\pmb{\Sigma}^{t}_{\infty})^{\frac{1}{2}},
\end{align}
where $\mathbf{d}^{t}_{n,\infty}$ is the corrupting noise in the converged matched filtered output.
At the same time, the Markov chain based activity model reveals the correlation between $\mathbf{x}^{t}_{n}$ and $(\mathbf{x}^{t-1}_{n},\mathbf{x}^{t+1}_{n})$, thus we can establish the correlation between $\mathbf{x}^{t}_{n}$ and $(\mathbf{r}^{t-1}_{n,\infty},\mathbf{r}^{t+1}_{n,\infty})$.
Note that the SI in \cite{Wang_2021_ISIT} is only extracted from $\mathbf{x}^{t}_{n}$ and $\mathbf{r}^{t-1}_{n,\infty}$.
In this work, we extract SI from the correlation among $\mathbf{x}^{t}_{n}$, $\mathbf{r}^{t-1}_{n,\infty}$ and $\mathbf{r}^{t+1}_{n,\infty}$. Specifically, the estimation results of $\mathbf{r}^{t-1}_{n,\infty}$ and $\mathbf{r}^{t+1}_{n,\infty}$ are used as the SI for device $n$, denoted as $\mathcal{S}^{t}_{n}=\{\mathbf{r}^{t-1}_{n,\infty},\mathbf{r}^{t+1}_{n,\infty} \}$, in the vAMP-SI framework\footnote{Note that we only consider one previous frame and one next frame for SI since the temporal correlation of device activity is modeled as a first-order Markov chain. An arbitrary number of adjacent frames can be considered if the device activity evolves according to an arbitrary stochastic process.}.
\begin{comment}
\begin{align}\label{equ:LLR_t}
    \upsilon^t_{n,\infty} =&~ \frac{\text{Pr}(\mathbf{r}^t_{n,\infty}|\lambda^t_n = 0)}{\text{Pr}(\mathbf{r}^t_{n,\infty}|\lambda^t_n = 1)} \notag \\
    %\quad &= \left(  \frac{\sigma^t_{\infty} + \beta}{\sigma^t_{\infty}} \right)^M \exp\left( - \Xi^t_{n,\infty} ||\mathbf{r}^t_{n,\infty}||_2^2 \right),
    \quad =&~ \frac{|\beta\mathbf{I}+\pmb{\Sigma}^{t}_{\infty}|}{|\pmb{\Sigma}^{t}_{\infty}|} \notag \\
    \quad &\times \exp (-\mathbf{r}^{t}_{n,\infty}((\pmb{\Sigma}^{t}_{\infty})^{-1}-(\beta\mathbf{I}+\pmb{\Sigma}^{t}_{\infty})^{-1})\mathbf{r}^{t}_{n,\infty}).
\end{align}
If the perfect SI from the $t$th frame is available, it means that the exact activity in the $t$th frame is known. In this case, we define $\upsilon^{t}_{n,\infty} = \infty$ if $\lambda^{t}_{n} = 0$, and $\upsilon^{t}_{n,\infty} = 0$ otherwise.
\end{comment}
In the following, we elaborate how to utilize the extracted SI for joint activity detection and channel estimation.
%will show the design of the shrinkage function $\pmb{\eta}_{n,i}(\mathbf{r}^{t}_{n,i},\mathcal{S}^{t}_{n})$ in the vAMP-SI framework, which takes advantage of the correlation among $\mathbf{x}^{t}_{n} , \mathbf{r}^{t-1}_{n,\infty}$ and $\mathbf{r}^{t+1}_{n,\infty}$.

\section{Double-Sided Information Aided Joint Activity Detection and Channel Estimation}\label{sec:GSI-JADCE}

%In this section, we introduce the joint activity detection and channel estimation algorithm under the vAMP-SI framework, which can take advantage of the double-sided information for performance improvement.

\subsection{Sliding-Window Detection}

\begin{figure}
    \centering
    \includegraphics[width=.4\textwidth]{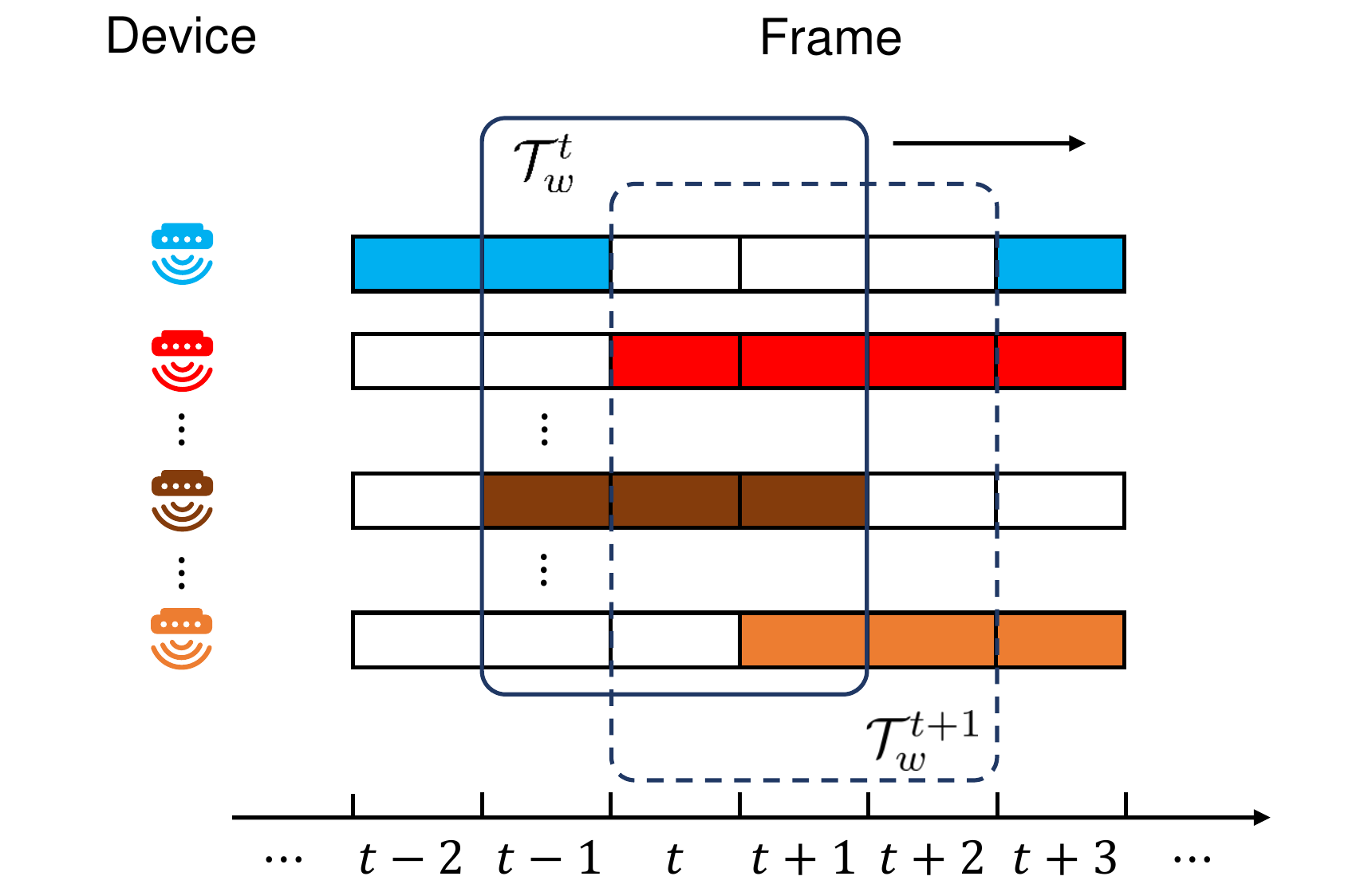}
    \vspace{-0.4cm}
    \caption{The diagram of the sliding-window detection strategy}
    \label{Fig:SW}
    \vspace{-0.5cm}
\end{figure}

%Note that the SI in \cite{Wang_2021_ISIT} only contains the estimation results in one single previous frame to improve the performance.
%Such strategy can only get a sub-optimal solution, which can not fully exploit the temporal correlation in activity.
%Therefore, in this work, we propose to utilize the SI from the estimation result in the previous frame and the next frame.
As shown in Fig. \ref{Fig:SW}, we perform joint activity detection and channel estimation in a sliding-window manner.
The sliding window is defined as $\mathcal{T}^{t}_{w} = \{t-1,t,t+1\}$. It includes the target detection frame and the frames from which the SI is acquired. After finishing the detection and estimation in frame $t$, the window slides to $\mathcal{T}^{t+1}_{w}$ to perform the task in the next frame.
%The window length can be obtained as $T_w = T_l + T_r + 1$ and there are totally $2^{T_w}$ activity patterns. For simplicity,
In each sliding window, there are totally $2^{3}$ possible activity patterns for each device. We denote each pattern as $\textbf{P}_s = \lambda^{t-1}_{n}\lambda^{t}_{n}\lambda^{t+1}_{n}$, and its occurrence probability as $u_s = p(\textbf{P}_s)$ for $s = 1,\dots,8$. The occurrence probabilities can be obtained by using the first-order Markov chain based temporal correlation as modeled in Section \ref{subsec:TC_Activity}.
For instance, we have $u_s = (1-p_a)(1-p_{01})^2$ when $\textbf{P}_s=000$ and $u_s = p_a p_{11}^2$ when $\textbf{P}_s=111$.

\vspace{-0.5cm}
\subsection{Shrinkage Function Design}\label{Subsec:SFD}

Assuming that the system statistics are available, we can design the shrinkage function based on the minimum mean-squared error (MMSE) criterion, known as the MMSE-optimal denoising function, to achieve the Bayesian optimality.
We first model the effective channel vectors $\mathbf{x}^{t}_{n}$ of each user $n$ in the sliding window $\mathcal{T}^{t}_{w}$ with a multivariable Bernoulli Gaussian distribution.
The probability distribution can be expressed as
\begin{align}\label{equ:MBG}
    p\left(\{\mathbf{x}^{\tau}_{n}\}_{\tau \in \mathcal{T}_w}\right) &= \sum_{s=1}^{8}\Bigg[ u_s \cdot \Bigg(\prod_{\tau=t-1}^{t+1} d_s(\mathbf{x}^{\tau}_{n}) \Bigg) \Bigg],
\end{align}
where
\begin{align}\label{equ:ds}
    d_s(\mathbf{x}^{\tau}_{n}) &= \left\{ \begin{array}{ll}
                                           \delta(\mathbf{x}^{\tau}_{n}), & \text{if}~ \lambda^{\tau}_n = 0 ~\text{in}~ \textbf{P}_s, \\
                                           \mathcal{CN}(\mathbf{x}^{\tau}_{n}; 0, \beta), & \text{if}~ \lambda^{\tau}_n =1 ~\text{in}~ \textbf{P}_s,
                                         \end{array} \right.
\end{align}
and we define $\mathcal{CN}(\mathbf{x}^{\tau}_{n}; 0, \beta) = (\pi\beta)^{-M}\exp(-\beta^{-1}||\mathbf{x}^{\tau}_{n}||^2_2)$.

Under the above probabilistic model, the MMSE-optimal denoising function is given as
\begin{align}\label{equ:MMSE_gen}
    \pmb{\eta}_{n,i}\left(\mathbf{r}^t_{n,i}, \mathcal{S}^{t}_{n} \right) &= \mathbb{E}\left[ \mathbf{x}^t_{n} | \mathbf{r}^t_{n,i}, \mathcal{S}^{t}_{n} \right].
\end{align}
%where $\mathcal{S}^{t}_{n} = \{\mathbf{r}^{\tau}_{n,\infty}\}_{\tau \in \mathcal{T}^t_w \setminus t}$ is the set containing the SI identified in Section \ref{subsec:SI_Ac}.
By substituting (\ref{equ:MMSE_gen}) into (\ref{equ:SE}), the state $\pmb{\Sigma}^{t}_{i}$ in each iteration will also be a scaled identity matrix as $\pmb{\Sigma}^t_{i} = e^{t}_{i}\mathbf{I}, \forall i,t$, if the initial state is $\pmb{\Sigma}^t_{0} = e^{t}_{0}\mathbf{I}, \forall i,t$. The proof is similar to that in \cite{Chen_2018_TSP,Liu_2018_TSP}.
With the above, the explicit expression of the MMSE-optimal denoising function can be given by the following Theorem.

\begin{theorem}\label{Thm:Thm_MMSE_denoiser1}
The MMSE-optimal denoising function for user $n$ in the vAMP framework with DSI can be represented as
\begin{align}\label{equ:MMSE_denoiser_gen}
    \pmb{\eta}_{n,i}\left(\mathbf{r}^t_{n,i},\mathcal{S}^{t}_{n}\right) =&~ %\frac{\left(1+\tau^t_{i} / \beta\right)^{-1}\widehat{\mathbf{r}}^t_{n,i}}{1 + \frac{\sum_{s \in \bar{\mathcal{S}}_w} p(\widehat{\mathbf{r}}^t_{n,i},\mathcal{R}^t_n,\textbf{P}_s)}{\sum_{s \in \mathcal{S}_w} p(\widehat{\mathbf{r}}^t_{n,i},\mathcal{R}^t_n,\textbf{P}_s)}}, \notag \\
    %\quad &=
    \frac{\left(1+e^t_{i} / \beta\right)^{-1}\mathbf{r}^t_{n,i}}{1+ \phi^{t}_{n,i}(\mathbf{r}^{t}_{n,i},\mathcal{S}^{t}_{n})},
    %\frac{\sum_{s \in \bar{\mathcal{S}}_w} u_{w,s} \varphi(\mathbf{r}^t_{n,i}, \textbf{P}_s) \prod_{t' \in \mathcal{T}^t_w}\varphi(\mathbf{r}^{t'}_{n,\infty}, \textbf{P}_s)  }{\sum_{s \in \mathcal{S}_w} u_{w,s} \varphi(\mathbf{r}^t_{n,i}, \textbf{P}_s) \prod_{t' \in \mathcal{T}^t_w}\varphi(\mathbf{r}^{t'}_{n,\infty}, \textbf{P}_s)}},
\end{align}
where
\begin{align}
    \phi^t_{n,i}(\mathbf{r}^{t}_{n,i},\mathcal{S}^{t}_{n}) =&~ \upsilon^t_{n,i} \times \frac{1-p_a}{p_a} \times \frac{p_{01}+(1-p_{01})\upsilon^{t-1}_{n,\infty}}{p_{11}+(1-p_{11})\upsilon^{t-1}_{n,\infty}} \notag \\
    \quad &\times \frac{p_{01}+(1-p_{01})\upsilon^{t+1}_{n,\infty}}{p_{11}+(1-p_{11})\upsilon^{t+1}_{n,\infty}}, \label{equ:func_phi}
\end{align}
and $\upsilon^t_{n,i}$ is the inverse LLR defined as
\begin{align}
    \upsilon^t_{n,i} = \frac{p(\mathbf{r}^{t}_{n,i}|\lambda^{t}_{n}=0)}{p(\mathbf{r}^{t}_{n,i}|\lambda^{t}_{n}=1)} = \left(\frac{\beta+e^{t}_{i}}{e^{t}_{i}}\right)^M\exp\Big(-\Xi^{t}_{i}||\mathbf{r}^{t}_{n,i}||^2_2\Big), \label{equ:inverse_LLR}
\end{align}
with $\Xi^{t}_{i} = (e^{t}_{i})^{-1} - (e^{t}_{i}+\beta)^{-1}$. The inverse LLRs $\upsilon^{t-1}_{n,\infty}$ and $\upsilon^{t+1}_{n,\infty}$ are similarly calculated by following (\ref{equ:inverse_LLR}).
\end{theorem}

\begin{IEEEproof}
Please see Appendix \ref{App:PF_T1}.
\end{IEEEproof}
%As shown in (\ref{equ:func_phi}), we can first calculate the inverse LLRs $\upsilon^{t-1}_{n,\infty}$ and $\upsilon^{t+1}_{n,\infty}$ and then utilize them to design the MMSE-optimal denoising function (\ref{equ:MMSE_denoiser_gen}) for helping estimate the effective channel vector $\mathbf{x}^{t}_{n}$ in the current frame.

To gain insights from Theorem \ref{Thm:Thm_MMSE_denoiser1}, we consider some special cases. First, when there is no temporal correlation in user activity, i.e., $p_{01} = p_{11} = p_a$, the function (\ref{equ:func_phi}) is simplified as $\phi^t_{n,i}(\mathbf{r}^{t}_{n,i},\mathcal{S}^{t}_{n}) = \frac{1-p_a}{p_a}\upsilon^t_{n,i}$. Thus, the function in (\ref{equ:MMSE_denoiser_gen}) reduces to the MMSE denoiser $\pmb{\eta}_{n,i}\left(\mathbf{r}^t_{n,i}\right) = \frac{\left(1+e^t_{i}/\beta\right)^{-1}\mathbf{r}^t_{n,i}}{1+\upsilon^{t}_{n,i}(1-p_a)/p_a}$ in \cite{Chen_2018_TSP, Liu_2018_TSP} without SI.
Second, if we have $\upsilon^{\tau}_{n,\infty} = 1$ for $\tau = t-1 \text{ or } t+1$, i.e., the device is equally likely to be active or inactive at frame $\tau$, then the SI from frame $\tau$ provides no useful information for the estimation.

To get more insights, we consider the extreme case where the exact activities in the previous frame and the next frame are known, i.e., the corresponding SI is \emph{perfect}. In this case, we have $\upsilon^{\tau}_{n,\infty} = \infty$ if $\lambda^{\tau}_{n} = 0$, and $\upsilon^{\tau}_{n,\infty} = 0$ otherwise, for $\tau \in \{t-1,t+1\}$.
With perfect SI, then when user activity is fully correlated in time with  $p_{11} = 1$, we have $\lambda^{t}_{n} = \lambda_n\in\{0,1\}$, $\forall t$.
%, then $\lambda^{t-1}_{n} = \lambda^{t}_{n} = \lambda^{t+1}_{n}$.
The MMSE-optimal denoising function becomes $ \pmb{\eta}_{n,i}(\mathbf{r}^{t}_{n,i},\mathcal{S}^{t}_{n}) = (1+e^{t}_{i}/\beta)^{-1}\mathbf{r}^{t}_{n,i}$ if $\lambda_{n} = 1$, or $\pmb{\eta}_{n,i}(\mathbf{r}^{t}_{n,i},\mathcal{S}^{t}_{n}) = \mathbf{0}$ if $\lambda_n=0$. As such, the joint activity detection and channel estimation reduces to channel estimation only, and the channel estimator is a linear function that operates on each element of $\mathbf{r}^{t}_{n,i}$ individually. This suggests that the performance of the algorithm is independent of the number of antennas.
%If $\upsilon^{t-1}_{n,\infty} = 1$ and $\upsilon^{t+1}_{n,\infty} = 1$, the MMSE-optimal denoising function in (\ref{equ:MMSE_denoiser_gen}) will also be reduced to that in \cite{Chen_2018_TSP, Liu_2018_TSP}.

%\subsection{Algorithm Design}
When detecting the active users in the $t$th frame, the estimation results in the $(t-1)$th frame is already available and thus the SI $\{\upsilon^{t-1}_{n,\infty}\}_{n=1}^{N}$ can be readily obtained.
On the other hand, the SI from the $(t+1)$th frame need to be estimated. In this work, we propose to employ the vAMP algorithm \cite{Chen_2018_TSP, Liu_2018_TSP} to perform coarse detection and estimation in the $(t+1)$th frame,
%\footnote{We can also use the SI-aided MMV-AMP algorithm \cite{Wang_2021_ISIT} to perform the coarse detection and estimation for the $(t+1)$th frame, however, the simulation trials show that this setting can provides only a little performance improvement but increase much complexity},
then the estimation results of $\{\mathbf{r}^{t+1}_{n,\infty}\}_{n=1}^{N}$ and $e^{t+1}_{\infty}$ are used to obtain $\{\upsilon^{t+1}_{n,\infty}\}^N_{n=1}$. Finally, we perform joint activity detection and channel estimation in the $t$th frame based on the vAMP-SI algorithm with $\{\upsilon^{t-1}_{n,\infty}\}^N_{n=1}$ and $\{\upsilon^{t+1}_{n,\infty}\}^N_{n=1}$.

%\begin{algorithm}[t]
%\renewcommand{\algorithmicrequire}{\textbf{Initialize:}}
%\renewcommand\algorithmicensure {\textbf{Output:} }
%\caption{The proposed detection procedure for the $t$th frame based on vAMP-SI}\label{Alg:vAMP-SI}
%\begin{algorithmic}[1]
%\STATE Initialize: $\{\upsilon^{t-1}_{n,\infty}\}^{N}_{n=1}$, $\mathbf{Y}^t$, $\mathbf{Y}^{t+1}$, $\mathbf{A}$.
%%\STATE Learn $\{\mathbf{W}\}$  based on the loss function $\mathcal{L}(\mathbf{\widehat{R}}_1) = ||\mathbf{\widehat{R}}_1 - \mathbf{X}^0||_F^2$.
%\STATE vAMP-SI1: Obtain $\widehat{\mathbf{X}}^{t+1}_{\infty}$ and $\{\upsilon^{t+1}_{n,\infty}\}^{N}_{n=1}$.
%\STATE vAMP-SI2: Obtain $\tilde{\mathbf{X}}^{t}_{\infty}$ and $\{\tilde{\upsilon}^{t}_{n,\infty}\}^{N}_{n=1}$ based on SI-aided AMP-MMV with $\{\upsilon^{t-1}_{n,\infty}\}^{N}_{n=1}$, then obtain $\widehat{\mathbf{X}}^{t+1}_{\infty}$ and $\{\upsilon^{t+1}_{n,\infty}\}^{N}_{n=1}$ by SI-aided AMP-MMV with $\{\tilde{\upsilon}^{t}_{n,\infty}\}^{N}_{n=1}$.
%\STATE Perform vAMP-SI to estimate $\widehat{\mathbf{X}}^t_{\infty}$ and $\{\upsilon^{t}_{n,\infty}\}^{N}_{n=1}$ with $\{\upsilon^{t-1}_{n,\infty}\}^{N}_{n=1}$ and $\{\upsilon^{t+1}_{n,\infty}\}^{N}_{n=1}$.
%\STATE Return $\widehat{\mathbf{X}}^t_{\infty}$ and $\{\upsilon^{t}_{n,\infty}\}^{N}_{n=1}$.
%\end{algorithmic}
%\end{algorithm}

%\vspace{-0.5cm}
\subsection{Activity Detector Design}

The activity detection is performed based on the LLR test after the proposed algorithm converges.
We denote $H_0$ as the hypothesis that the device $n$ is inactive with $\lambda^t_n = 0$, and denote $H_1$ as the hypothesis otherwise. The Bayes-Risk activity decision rule is represented as
\begin{align}\label{equ:DR}
    \xi^t_{n} = \log\left( \frac{p(\mathbf{r}^t_{n,\infty},\mathcal{S}^{t}_{n}|\lambda^t_n = 1)}{p(\mathbf{r}^t_{n,\infty},\mathcal{S}^{t}_{n}|\lambda^t_n = 0)} \right) \mathop{\gtrless}\limits_{H_0}^{H_1} \iota^{\xi},
\end{align}
where $\xi^t_{n}$ is the LLR for device $n$ and $\iota^{\xi}$ is a predetermined decision threshold for all devices. Since the LLR $\xi^t_{n}$ is a monotonic function on $||\mathbf{r}^t_{n,\infty}||^2_2$, the decision rule can be simplified as $||\mathbf{r}^t_{n,\infty}||^2_2 \mathop{\gtrless}\limits_{H_0}^{H_1} \iota^{r}_{n}$.

\begin{theorem}\label{Thm:GDT}
The decision threshold of user $n$ in the $t$th frame can  be expressed as
\begin{small}
\begin{align}\label{equ:GDT}
    \iota^{r}_{n} &= \frac{\iota^{\xi} + M\log\Big( \frac{e^t_{\infty}+\beta}{e^t_{\infty}} \Big) + \log\Big(\frac{p_a}{1-p_a}\cdot\frac{\phi^{t}_{n,\infty}(\mathbf{r}^{t}_{n,\infty}, \mathcal{S}^{t}_{n})}{\upsilon^t_{n,\infty}}\Big)}{\Xi^t_{\infty}}.
    %\notag \\
    %\quad &~+ \frac{\log\Big(\frac{p_a}{1-p_a}\Big)}
    %\frac{\sum_{s \in \bar{\mathcal{S}}_w} u_{w,s} \prod_{t' \in \mathcal{T}^t_w}\varphi(\mathbf{r}^{t'}_{n,\infty}, \textbf{P}_s)  }{\sum_{s \in \mathcal{S}_w} u_{w,s} \prod_{t' \in \mathcal{T}^t_w}\varphi(\mathbf{r}^{t'}_{n,\infty}, \textbf{P}_s)}\Big)}{\Delta^t_{n,i}}.
\end{align}
%\begin{align}\label{equ:DT}
%    \iota^{r}_{n} =&~ \frac{\iota^{\xi}_{n} + M\log\Big( \frac{\sigma^t_{i}+\beta}{\sigma^t_{i}} \Big)}{\Xi^t_{n,\infty}} \notag \\
%    \quad &+ \frac{\log\Big(\frac{p_{01}+(1-p_{01})\upsilon^{t-1}_{n,\infty}}{p_{11}+(1-p_{11})\upsilon^{t-1}_{n,\infty}} \times \frac{p_{01}+(1-p_{01})\upsilon^{t+1}_{n,\infty}}{p_{11}+(1-p_{11})\upsilon^{t+1}_{n,\infty}}\Big)}{\Xi^t_{n,\infty}}.
%\end{align}
\end{small}where $\phi^{t}_{n,\infty}(\mathbf{r}^{t}_{n,\infty}, \mathcal{S}^{t}_{n})$ is given by (\ref{equ:func_phi}).

\end{theorem}

\begin{IEEEproof}
The proof is similar to that of Theorem \ref{Thm:Thm_MMSE_denoiser1} and hence ignored.
\end{IEEEproof}

%In our system setting with $\mathcal{T}_w = \{t-1, t+1\}$, the decision threshold can be further simplified as
%\begin{align}\label{equ:DT}
%    \iota^{r}_{n} =&~ \frac{\iota^{\xi}_{n} + M\log\Big( \frac{\sigma^t_{i}+\beta}{\sigma^t_{i}} \Big)}{\Delta^t_{n,i}} \notag \\
%    \quad &+ \frac{\log\Big(\frac{p_{01}+(1-p_{01})\upsilon^{t-1}_{n,\infty}}{p_{11}+(1-p_{11})\upsilon^{t-1}_{n,\infty}} \times \frac{p_{01}+(1-p_{01})\upsilon^{t+1}_{n,\infty}}{p_{11}+(1-p_{11})\upsilon^{t+1}_{n,\infty}}\Big)}{\Delta^t_{n,i}}
%\end{align}

%With the activity detector, the device $n$ is detected to be active in the $t$th frame if $||\mathbf{r}^t_{n,\infty}||^2_2 > \iota^{r}_{n,i}$.
Theorem \ref{Thm:GDT} indicates that the SI from the previous and next frames has a great impact on the decision threshold setting. Similar to the insights derived from Theorem \ref{Thm:Thm_MMSE_denoiser1}, if $p_{01} = p_{11} = p_a$, the activity detector will be reduced to that in \cite{Chen_2018_TSP, Liu_2018_TSP}, where the SI is not utilized. When we have $\upsilon^{\tau}_{n,\infty} = 1$ for $\tau = t-1 \text{ or } t+1$, no useful information is provided from the $\tau$th frame for activity detection.
%In the extreme case where the perfect SI is available and $p_{11} = 1$, we will have $\iota^{r}_{n} = \infty$ if $\lambda_{n} = 0$ and $\iota^{r}_{n} = -\infty$ otherwise. Therefore, the activity can be directly detected as $\widehat{\lambda}^{t}_{n} = \lambda^{t-1}_{n} = \lambda^{t+1}_{n}$, which is consistent to the result of deciding the activity from the probability perspective with $p(\lambda^{t}_{n}=1) = p(\lambda^{t-1}_{n}=1)$.

\subsection{Discussion}
Compared with the SI-aided MMV-AMP algorithm \cite{Wang_2021_ISIT} that only exploits the single-sided information (SSI), our proposed algorithm can utilize the DSI. 
In particular, the SI from frame $(t+1)$ is obtained by performing an extra vAMP estimation on $\mathbf{X}^{t+1}$. This, however, doubles the computation cost and adds one-frame detection delay.
%Therefore, if higher computational complexity and the one-frame detection delay can be tolerated, significant performance enhancement will be provided by our proposed algorithm.
As shall be demonstrated in the next section, such extra complexity and delay can bring significant performance enhancement when the BS only has a few antennas. 

%\begin{comment}

\subsection{Extension to the Generalized Case}

Here, we extend the proposed algorithm to the generalized case where the activity evolution of the device follows an arbitrary stochastic process. Since the activity in this case may not be only correlated with those in the previous frame and the next frame, we can consider a larger sliding window starting from the $(t-T_l)$th frame to the $(t+T_r)$th frame with $\mathcal{T}^t_w = \{\tau|\tau = t-T_l, \dots, t+T_r, \tau \ne t\}$. With the prior information of the activity evolution and the SI from the estimation results of all frames in $\mathcal{T}^t_w$, the MMSE-optimal denoising function and the decision threshold for each user $n$ can will be given by the following corollary.

\begin{corollary}\label{Cor:MMSE_generalized}
In the generalized case where the activity evolution of the device follows an arbitrary stochastic process, the MMSE-optimal denoising function and the decision threshold for each user $n$ in the $t$th frame are also written in the form of (\ref{equ:MMSE_denoiser_gen}) and (\ref{equ:GDT}). While the function $\phi^t_{n,i}(\mathbf{r}^{t}_{n,i},\mathbf{s}^{t}_{n})$ is modified as
\begin{align}\label{equ:phi}
    \phi^{t}_{n,i}(\mathbf{r}^{t}_{n,i},\mathbf{s}^{t}_{n}) &= \frac{\sum_{s \in \bar{\mathcal{S}}_w} u_{s} \varphi^{t}_{i}(\mathbf{r}^t_{n,i}, \textbf{P}_s) \prod_{\tau \in \mathcal{T}^t_w}\varphi^{\tau}_{\infty}(\mathbf{r}^{\tau}_{n,\infty}, \textbf{P}_s)  }{\sum_{s \in \mathcal{S}_w} u_{s} \varphi^{t}_{i}(\mathbf{r}^t_{n,i}, \textbf{P}_s) \prod_{\tau \in \mathcal{T}^t_w}\varphi^{\tau}_{\infty}(\mathbf{r}^{\tau}_{n,\infty}, \textbf{P}_s)}.
\end{align}
where $\varphi^{t}_{i}(\mathbf{r}^t_{n,i}, \textbf{P}_s)$ and $\varphi^{\tau}_{\infty}(\mathbf{r}^{\tau}_{n,\infty}, \textbf{P}_s)$ are defined in (\ref{equ:func_varphi}).
\end{corollary}

\begin{IEEEproof}
The proof of this corollary is similar to the proof of Theorem \ref{Thm:Thm_MMSE_denoiser1} and hence ignored.
\end{IEEEproof}

Based on the corollary, we can see that the MMSE-optimal denoising function can be designed in the explicit expression once the occurrence probability $u_s$ is available even if the activity evolution cannot be explicitly characterized.

%\end{comment}

%\begin{align}\label{equ:PDR}
%    \iota^{r}_{n,i} &= \frac{\iota^{\xi}_{n} + M\log\Big( \frac{\tau^t_{i}+\beta}{\tau^t_{i}} \Big) + \log\Big( \frac{p_a}{1-p_a}\phi^t_{n,i}(\widehat{\mathbf{r}}^{t-1}_{n,\infty},\widehat{\mathbf{r}}^{t+1}_{n,\infty}) \Big)}{\Delta^t_{n,i}}.
%\end{align}

%For the generalized case where the detection window contains $l_w$ frames and starts from the $(t-l_l)$th frame to the $(t+l_r)$th frame, the decision threshold can be given with the following lemma.

%\begin{lemma}
%Under the proposed detection and estimation strategy, the decision threshold of user $n$ in the $t$th frame for the generalized case can be expressed as
%\begin{align}\label{equ:GDT}
%    \iota^{r}_{n,i} =&~ \frac{\iota^{\xi}_{n} + M\log\Big( \frac{\tau^t_{i}+\beta}{\tau^t_{i}} \Big)}{\Delta^t_{n,i}} \notag \\
%    \quad &~+ \frac{\log\Big(\frac{p_a}{1-p_a}\frac{\sum_{s \in \bar{\mathcal{S}}_w} u_{w,s} \prod_{t' \in \mathcal{T}^t_w}\varphi(\mathbf{r}^{t'}_{n,\infty}, \textbf{P}_s)  }{\sum_{s \in \mathcal{S}_w} u_{w,s} \prod_{t' \in \mathcal{T}^t_w}\varphi(\mathbf{r}^{t'}_{n,\infty}, \textbf{P}_s)}\Big)}{\Delta^t_{n,i}}.
%\end{align}
%\end{lemma}
%
%\begin{IEEEproof}
%This lemma can be similarly obtained by following the acquisition of (\ref{equ:PDR}).
%\end{IEEEproof}

\section{Simulation Results}

\begin{figure}[t]
  \centering
  \includegraphics[width=.4\textwidth]{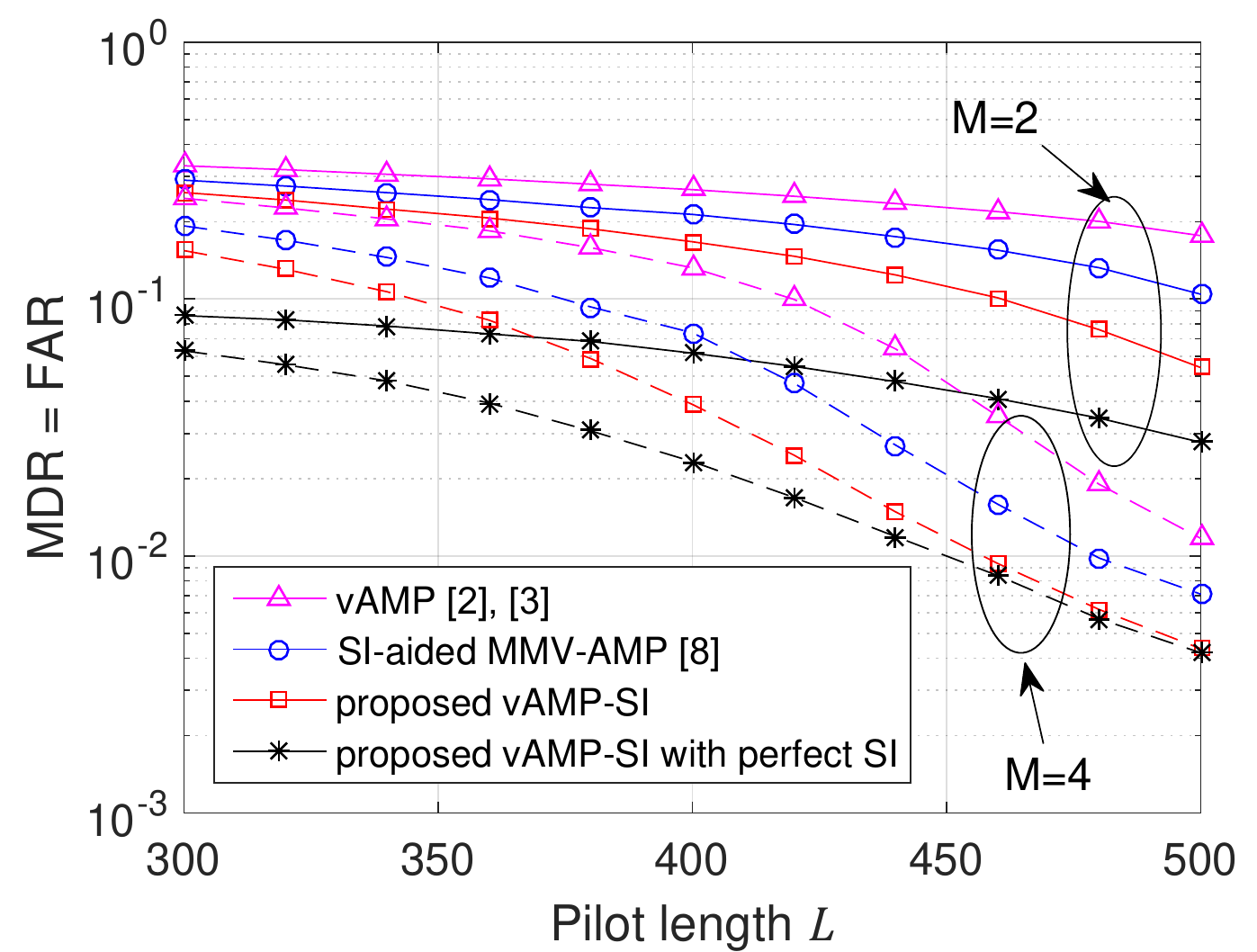}
  \vspace{-0.4cm}
  \caption{Activity detection performance versus pilot length $L$.}\label{Fig:ER_L}
  \vspace{-0.2cm}
\end{figure}

\begin{figure}[t]
  \centering
  \includegraphics[width=.4\textwidth]{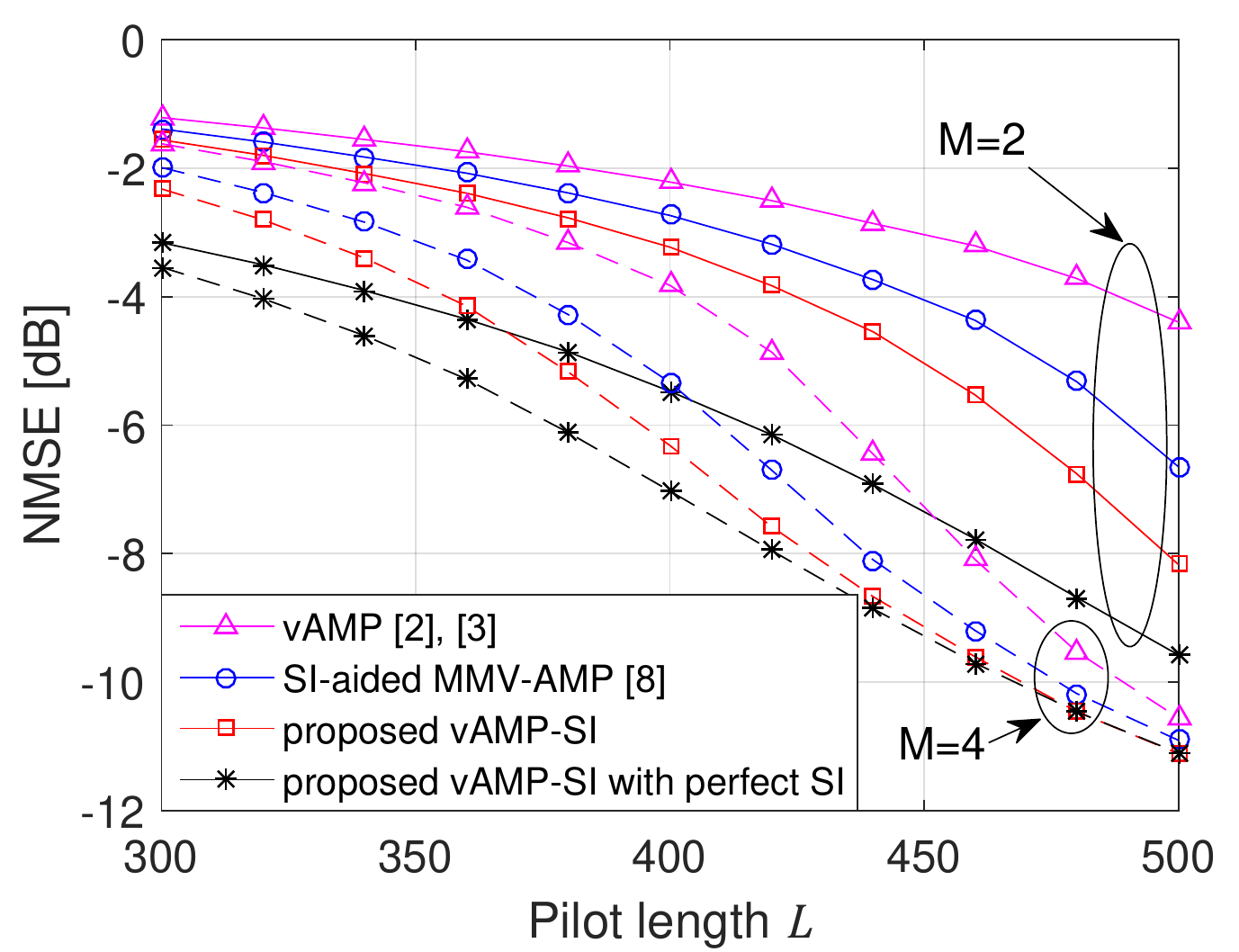}
  \vspace{-0.4cm}
  \caption{Channel estimation performance versus pilot length $L$.}\label{Fig:NMSE_L}
  \vspace{-0.4cm}
\end{figure}

\begin{figure}[t]
  \centering
  \includegraphics[width=.4\textwidth]{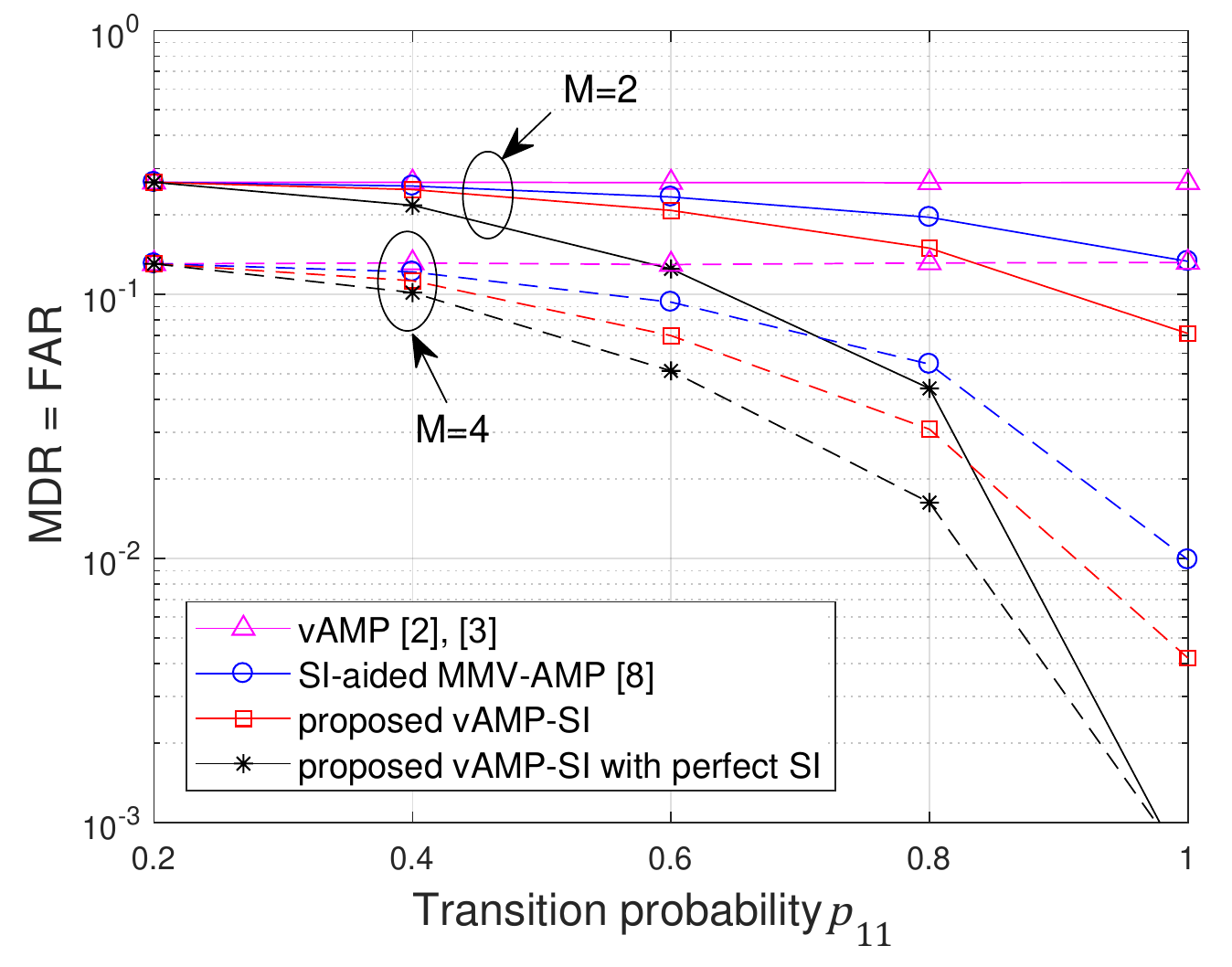}
  \vspace{-0.5cm}
  \caption{Activity detection performance versus $p_{11}$.}\label{Fig:ER_p11}
  \vspace{-0.2cm}
\end{figure}

\begin{figure}[t]
  \centering
  \includegraphics[width=.4\textwidth]{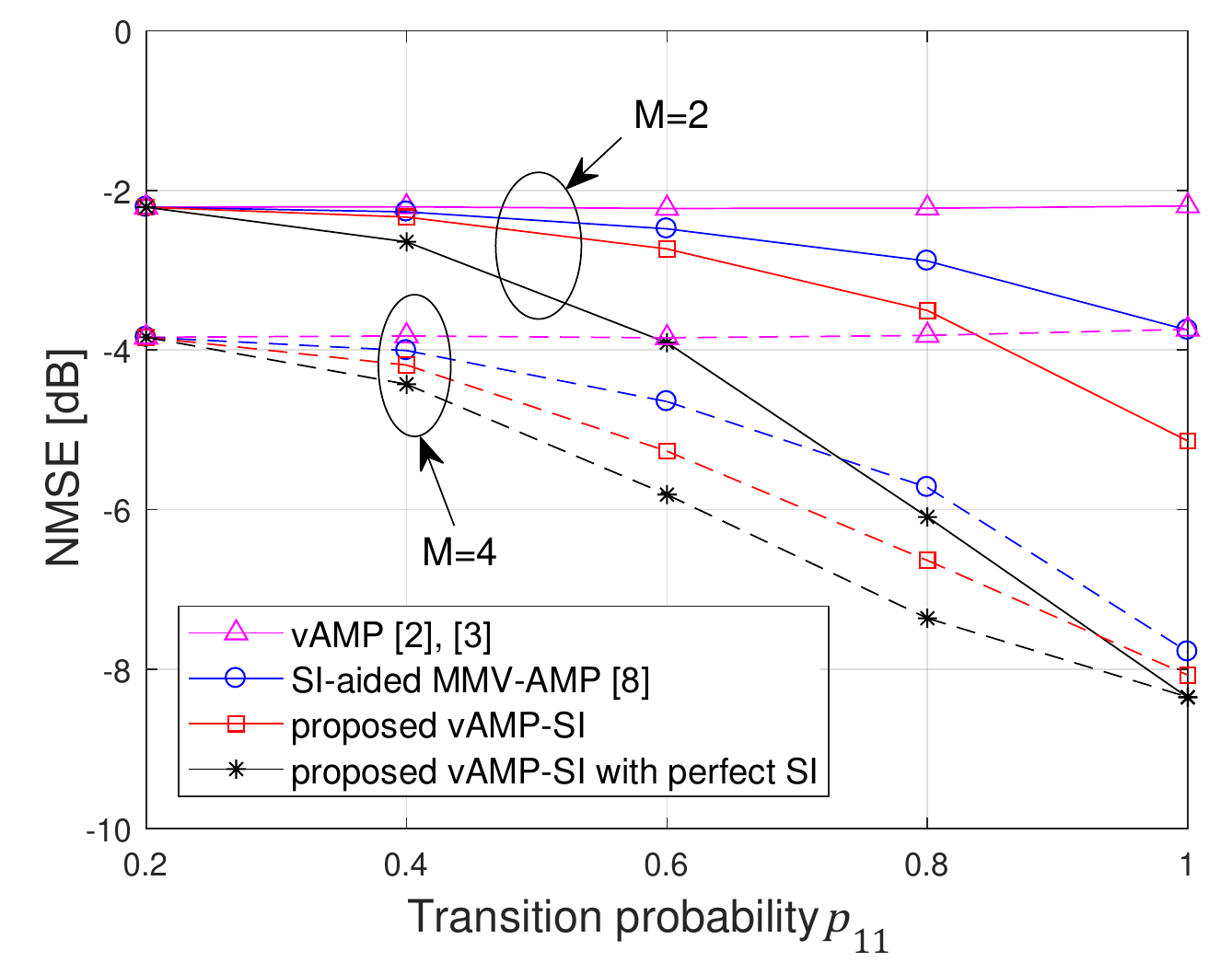}
  \vspace{-0.5cm}
  \caption{Channel estimation performance versus $p_{11}$.}\label{Fig:NMSE_p11}
  \vspace{-0.7cm}
\end{figure}

We consider the system containing $N=2000$ devices. The signal-to-noise ratio is defined as $\text{SNR} = \frac{\beta}{\sigma^2_w}$ and is set to $\text{SNR} = -10\text{dB}$. We set $p_{01} = 1/16$ and $p_{11} = 3/4$ if not specified otherwise, then we have $p_a = 1/5$. The joint activity detection and channel estimation is performed in $T=8$ consecutive frames. The missed detection ratio (MDR) and the false alarm ratio (FAR) are used as the performance metrics for activity detection, the normalized mean-squared error (NMSE) is used as the performance metric for channel estimation. The vAMP algorithm \cite{Liu_2018_TSP, Chen_2018_TSP} and the SI-aided MMV-AMP algorithm \cite{Wang_2021_ISIT} are considered as benchmarks. We also evaluate the proposed algorithm with perfect SI as performance bound.

%\begin{figure}[t]
%  \centering
%  \includegraphics[width=.45\textwidth]{Fig_ER_L.eps}
%  \caption{device activity detection performance comparison}\label{Fig:ER_L}
%\end{figure}
%
%\begin{figure}[htbp]
%  \centering
%  \includegraphics[width=.45\textwidth]{Fig_NMSE_L.eps}
%  \caption{Channel estimation performance comparison}\label{Fig:NMSE_L}
%\end{figure}

%\begin{figure}[t]
%  \centering
%  \begin{minipage}[t]{.24\textwidth}
%    \center
%    \includegraphics[width=\textwidth]{Fig_ER_L.pdf}
%    \vspace{-0.7cm}
%    \caption{Activity detection performance versus pilot length $L$}\label{Fig:ER_L}
%  \end{minipage}
%  \begin{minipage}[t]{.24\textwidth}
%    \center
%         \includegraphics[width=\textwidth]{Fig_NMSE_L.pdf}
%    \vspace{-0.7cm}
%    \caption{Channel estimation performance versus pilot length $L$}\label{Fig:NMSE_L}
%  \end{minipage}
%\vspace{-0.2cm}
%\end{figure}

Fig. \ref{Fig:ER_L} and Fig. \ref{Fig:NMSE_L} show the activity detection and the channel estimation performance, respectively, at different pilot lengths. Here, we have set $\text{MDR} = \text{FAR}$ by carefully selecting the decision threshold $ \iota^{r}_{n} $. We observe that the proposed algorithm can significantly outperform the state-of-the-art algorithms. We also observe that the performance gap to the lower bound with perfect SI reduces when the antenna number $M$ and the pilot length $L$ increases. In particular, when $M=4$ and $L=500$, the proposed vAMP-SI algorithm nearly converges to the ideal case. This indicates that the proposed algorithm can better exploit the temporal correlation in device activity to improve the performance.

%\begin{figure}[t]
%  \centering
%  \begin{minipage}[t]{.24\textwidth}
%    \center
%    \includegraphics[width=\textwidth]{Fig_ER_p11.pdf}
%    \vspace{-0.3cm}
%    \caption{Activity detection performance versus $p_{11}$}\label{Fig:ER_p11}
%  \end{minipage}
%  \begin{minipage}[t]{.24\textwidth}
%    \center
%         \includegraphics[width=\textwidth]{Fig_NMSE_p11.pdf}
%    \vspace{-0.3cm}
%    \caption{Channel estimation performance versus $p_{11}$}\label{Fig:NMSE_p11}
%  \end{minipage}
%\vspace{-0.2cm}
%\end{figure}

Fig. \ref{Fig:ER_p11} and Fig. \ref{Fig:NMSE_p11} illustrate the performance of the proposed algorithm under different transition probabilities $p_{11}$, where the active probability is fixed to be $p_a=1/5$. As mentioned, when $p_{01} = p_{11} = p_a$, the device activities in different frames are temporally independent. Thus, we consider that $p_{11}$ ranges from 0.2 to 1 in the simulations. It is observed that by exploiting the double-sided information, the proposed vAMP-SI algorithm outperforms considerably the existing SI-aided MMV-AMP. Similar to the result in Fig. \ref{Fig:ER_L} and Fig. \ref{Fig:NMSE_L}, the performance of the proposed algorithm is closer the ideal performance with perfect SI by increasing $M$. When $p_{11}=1$, the proposed algorithm with perfect SI has the same channel estimation performance for $M=2$ and $M=4$. This observation validates the discussion in Section \ref{Subsec:SFD} that the channel estimation performance in this limiting case is irrelevant to the number of antennas.

%\vspace{-0.2cm}
\section{Conclusion}

This paper proposes to extract the SI from the estimation results in both the previous frame and the next frame for joint activity detection and channel estimation in temporal-correlated massive access. We introduce a sliding-window detection strategy and develop the vAMP-SI algorithm to exploit the double-sided information.
%Then the MMSE-optimal denoising function and the activity detector under vAMP-SI are also given in the generalized case where the activity evolves to an arbitrary stochastic process.
%the MMSE-optimal denoising function and the generalized LLR activity detector are designed to account the SI from the any number of neighboring frames. Then
The numerical results have validated that the proposed algorithm can achieve superior performance over the start-of-the-art algorithms.
%\textcolor[rgb]{0.00,0.07,1.00}{This work only obtains the SI from the estimation results in both the previous and the next frames due to the first-order Markov activity model. For more generalized activity evolution model, the SI-aided AMP algorithm with a larger sliding window can be investigated. The channel may also evolves according to a stochastic process and can be exploited in the algorithm design.}
This work considers a block fading channel model, while the channel may also evolves according to a stochastic process and can be exploited in the algorithm design.

%\appendix

%\subsection{Proof of Theorem 1}\label{App:PF_T1}

\small

\begin{appendices}

\section{Proof of Theorem \ref{Thm:Thm_MMSE_denoiser1}}\label{App:PF_T1}

The MMSE-optimal denoising function function can be written as
\begin{small}
\begin{align}\label{equ:DF}
    \pmb{\eta}_{n,i}  \left(\mathbf{r}^t_{n,i},\mathcal{S}^{t}_{n}\right)  &\triangleq \mathbb{E}[\mathbf{x}^t_{n,i}|\mathbf{r}^t_{n,i},\mathcal{S}^{t}_{n}] \notag \\
    \quad &= \sum_{s \in \mathcal{P}_w} \mathbf{E}\left[ \mathbf{x}^t_{n,i} |\mathbf{r}^t_{n,i},\mathcal{S}^{t}_{n}, \textbf{P}_s \right] p(\textbf{P}_s|\mathbf{r}^t_{n,i},\mathcal{S}^{t}_{n}),
\end{align}
\end{small}where $\mathcal{P}_w = \{s|\lambda^{t}_{n}=1 \text{ in } \mathbf{P}_s\}$ is the set containing the indexes of all activity patterns where $\lambda^{t}_{n}=1$.
%where $\mathcal{S} \in \{ s | \lambda^t_n = 1 ~\text{in}~ \textbf{P}_s \}$ denotes the set containing the index of the scenario $\textbf{P}_s$ where $\lambda^t_n = 1$ in the activity pattern. To simplify the representation, we define a function $\varphi(\mathbf{r}^t_{n,i}, \textbf{P}_s)$ as
%\begin{small}
%\begin{equation}\label{equ:PDF_r}
%    \varphi(\mathbf{r}^t_{n,i}, \textbf{P}_s) = \left\{ \begin{array}{ll}
%                                           \mathcal{CN}(\mathbf{r}^t_{n,{i}}; 0, \tau^t_{i}), & \text{if}~ \lambda^t_n = 0 ~\text{in}~ \textbf{P}_s, \\
%                                           \mathcal{CN}(\mathbf{r}^t_{n,{i}}; 0, \tau^t_{i}+\beta), & \text{if}~ \lambda^t_n =1 ~\text{in}~ \textbf{P}_s.
%                                         \end{array} \right. \\
%\end{equation}
%\end{small}
For $s \in \mathcal{P}_w$, we can obtain
\begin{small}
\begin{align}\label{equ:Exs}
    \mathbf{E}\left[ \mathbf{x}^t_{n,i} |\mathbf{r}^t_{n,i},\mathcal{S}^{t}_{n}, \textbf{P}_s \right] &\overset{(a)}{=} \mathbb{E}[\mathbf{x}^t_{n,i}|\mathbf{r}^t_{n,i} = \mathbf{h}^t_{n} + \sqrt{e^t_{i}}\mathbf{d}^t_n] \notag \\
    \quad &= \left(1+e^t_{i} / \beta\right)^{-1}\mathbf{r}^t_{n,i},
\end{align}
\end{small}where $\overset{(a)}{=}$ is the fact that all vectors in $\{\mathbf{h}^{\tau}_{n}\}_{\tau = t-1}^{t+1}$ and $\{\mathbf{d}^{\tau}_{n}\}_{\tau=t-1}^{t+1}$ are mutually independent. %and $\mathbf{x}^t_n = \mathbf{h}^t_n$ in each scenario $\textbf{P}_s$.
Then the probability $p(\textbf{P}_s|\mathbf{r}^t_{n,i}, \mathcal{S}^{t}_{n})$ can be calculated as
\begin{small}
\begin{align}\label{equ:Pps}
%    \quad   &~ p(\textbf{P}_s|\widehat{\mathbf{r}}^t_{n,i},\widehat{\mathbf{r}}^{t-1}_{n,\infty},\widehat{\mathbf{r}}^{t+1}_{n,\infty}) \notag \\
%    \quad =&~ \frac{p(\widehat{\mathbf{r}}^t_{n,i},\widehat{\mathbf{r}}^{t-1}_{n,\infty},\widehat{\mathbf{r}}^{t+1}_{n,\infty}|\textbf{P}_s) p(\textbf{P}_s)}{p(\widehat{\mathbf{r}}^t_{n,i},\widehat{\mathbf{r}}^{t-1}_{n,\infty},\widehat{\mathbf{r}}^{t+1}_{n,\infty})} \notag \\
    p(\textbf{P}_s|\mathbf{r}^t_{n,i},\mathcal{S}^{t}_{n}) &= \frac{p(\mathbf{r}^t_{n,i},\mathcal{S}^{t}_{n}|\textbf{P}_s) p(\textbf{P}_s)}{p(\mathbf{r}^t_{n,i},\mathcal{S}^{t}_{n})} \notag \\
    \quad & = \frac{u_s \varphi^{t}_{i}(\mathbf{r}^t_{n,i}, \textbf{P}_s) \prod_{\tau\in\mathcal{T}_w \setminus t} \varphi^{\tau}_{\infty}(\mathbf{r}^{\tau}_{n,\infty}, \textbf{P}_s)}{p(\mathbf{r}^t_{n,i},\mathcal{S}^{t}_{n})},
\end{align}
\end{small}where
\begin{small}
\begin{align}
    \varphi^{t}_{i}(\mathbf{r}^t_{n,i}, \textbf{P}_s) &= \left\{ \begin{array}{ll}
                                           \mathcal{CN}(\mathbf{r}^t_{n,i}; 0, e^t_{i}), & \text{if}~ \lambda^t_n = 0 ~\text{in}~ \textbf{P}_s, \\
                                           \mathcal{CN}(\mathbf{r}^t_{n,i}; 0, e^t_{i}+\beta), & \text{otherwise},
                                         \end{array} \right. \label{equ:func_varphi}
\end{align}
\end{small}
\begin{small}
\begin{align}
  p(\mathbf{r}^t_{n,i},\mathcal{S}^{t}_{n}, \textbf{P}_s) &=  \sum_{s=1}^{8} p(\mathbf{r}^t_{n,i},\mathcal{S}^{t}_{n}|\textbf{P}_s) p(\textbf{P}_s) \notag \\
  \quad &= \sum_{s=1}^{8} \left[ u_s \varphi^{t}_{i}(\mathbf{r}^t_{n,i}, \textbf{P}_s) \Bigg(\prod_{\tau \in \mathcal{T}^t_w \setminus t}\varphi^{\tau}_{\infty}(\mathbf{r}^{\tau}_{n,\infty}, \textbf{P}_s)\Bigg)\right]. \label{equ:p_joint}
\end{align}
\end{small}By inserting the above functions (\ref{equ:Exs}), (\ref{equ:Pps}) and (\ref{equ:p_joint}) into (\ref{equ:DF}), the MMSE-optimal denoising function in the vAMP-SI algorithm can be finally simplified in the form of (\ref{equ:MMSE_denoiser_gen}).

\end{appendices}

\bibliographystyle{IEEEtran}
\bibliography{TCMC_MMV_ref}

\end{document}